\documentclass[conference]{IEEEtran}

\usepackage{cite}
\usepackage{color}
\usepackage{multirow}
\usepackage{booktabs,multirow,array}
\usepackage[table]{xcolor} 
\usepackage{ltxcmds}
\usepackage{footnote}
\usepackage{amsthm}

\usepackage[table]{xcolor}
\usepackage{subfigure}
\usepackage{lipsum}
\usepackage{amssymb} 
\usepackage{pdflscape}
\usepackage{afterpage}

\usepackage{balance}

\usepackage{pifont}
\newcommand{\xmark}{\ding{55}}%

\ifCLASSINFOpdf
\usepackage[pdftex]{graphicx}
\else
  \usepackage[dvips]{graphicx}
\fi

\begin{document}
\title{A Trusted, Verifiable and Differential Cyber Threat Intelligence Sharing Framework using Blockchain}

\author{\IEEEauthorblockN{Kealan Dunnett$^{\ast}$, Shantanu Pal$^{\ast\ast}$, Guntur Dharma Putra$^{\mathsection}$, Zahra Jadidi$^{\mathsection\mathsection}$, Raja Jurdak$^{\ast}$}
\IEEEauthorblockA{$^{\ast} $School of Computer Science, Queensland University of Technology, Brisbane, QLD 4000, Australia\\ $^{**}$School of Information Technology, Deakin University, Melbourne, VIC 3125, Australia \\ $^{\mathsection}$School of Computer Science and Engineering, University of New South Wales, Sydney, NSW 2052, Australia \\ $^{\mathsection\mathsection}$School of Information and Communication Technology, Griffith University, Gold Coast Campus, QLD 4222, Australia
\\
{kealan.dunnett@connect.qut.edu.au,
shantanu.pal@deakin.edu.au,  gdputra@unsw.edu.au, z.jadidi@griffith.edu.au} {}}}




\maketitle

\begin{abstract}

Cyber Threat Intelligence (CTI) is the knowledge of cyber and physical threats that help mitigate potential cyber attacks. The rapid evolution of the current threat landscape has seen many organisations share CTI to strengthen their security posture for mutual benefit. However, in many cases, CTI data contains  attributes (e.g., software versions) that have the potential to leak sensitive information or cause reputational damage to the sharing organisation. While current approaches allow restricting CTI sharing to trusted organisations, they lack solutions where the shared data can be verified and disseminated `differentially' (i.e., selective information sharing) with policies and metrics flexibly defined by an organisation. In this paper, we propose a  blockchain-based CTI sharing framework that allows organisations to share sensitive CTI data in a trusted, verifiable and differential manner. We discuss the limitations associated with existing approaches and highlight the advantages of the proposed CTI sharing framework. We further present a detailed proof of concept using the Ethereum blockchain network. Our experimental results show that the proposed framework can facilitate the exchange of CTI without creating significant additional overheads.
\end{abstract}


%
\IEEEpeerreviewmaketitle

\section{Introduction}
\label{introduction}

The rapid evolution of the threat landscape in recent years has caused many organisations to reassess the risks associated with their digital infrastructure \cite{eling2021cyber}. For example, during the 2020-2021 financial year, almost 500 ransomware cybercrimes were reported in Australia, an increase of 15\% compared to the previous year \cite {ACSC-CT-report-2020-21}. As a result, many organisations have started integrating Cyber Threat Intelligence (CTI) into traditional Cyber Security Risks Management (CSRM) frameworks to try and achieve a more proactive security posture \cite{kure2019cyber}. CTI can be defined as the collection and synthesis of evidence-based knowledge about actors, threats, and vulnerabilities of harmful events occurring in cyberspace. Subsequently, the contents of CTI are highly variable in nature and can range from Indicators of Compromise (IoC) to sophisticated information detailing an attacker's Techniques, Tactics and Procedures (TTPs) \cite{wagner2019cyber}. 

Given that integration of CTI data into traditional CSRM frameworks can be utilised to provide organisations with a more proactive defense posture, inter organisational CTI sharing holds additional benefits \cite{kure2019cyber}. In the case where one organisation (e.g., a CTI producer) has been attacked, sharing CTI has the potential to allow other organisations (e.g., a CTI consumer) to proactively implement mitigation strategies \cite{wagner2019cyber}. Note, we refer to CTI producers as the entities who share CTI, while CTI consumers are the entities that use shared CTI. As a result, in recent years several \textit{Open-source/Vendor-created} CTI sharing platforms have been developed. 

\subsection{Motivation and Problem Statement}
In many cases, sharing CTI data in an uncontrolled way can result in CTI producers sustaining reputational damage.
Subsequently, for cases where CTI contains sensitive data, trust-based data sharing has been highlighted as a possible solution \cite{wagner2019cyber}. Trust-based CTI sharing ensures that potentially sensitive CTI data is shared with consumers with which pre-established or developed trust exists. Moreover, trust issues are significant when sharing takes place in large-scale industrial systems, e.g., Industry 4.0~\cite{suhail2022towards}.

While trust-based sharing has the potential to reduce sharing risks, CTI producers require more fine-grained control over what data they share. For instance, in \cite{wagner2019cyber}, it is highlighted that organisations are likely to differ in their perceptions of what data is considered sensitive. 
As a result, CTI sharing platforms need to provide organisations with the flexibility to share CTI in a way that caters to this variation. Organisations are also likely to differ in their perception of how trust-based sharing should be facilitated. For instance, an organisation A may determine their trust in another organisation B based on B's attribute or using identity-based policies specific to B. Subsequently, constraining organisations to share in a fixed way, either through  share or deny or policies, or group-based policies that share or deny CTI with specific groups, does not provide the granularity required for versatile CTI sharing. We propose a framework that allows CTI producers to share sensitive CTI with consumers in a \textit{differential} way. We refer to differential as the ability of CTI producers to control the amount of information exchanged with CTI consumers at a fine-grained level (more discussion in Section~\ref{differential}).  

Furthermore, CTI consumers should have the ability to verify requested information to ensure that the supplied intelligence has not be tampered with or altered by the CTI producer. Subsequently, an efficient verification mechanism which simultaneously preserves the privacy in data sharing is required. 
Thus, we argue that CTI sharing must be trusted, performed in a verifiable way, and without compromising the privacy of sensitive CTI data.

Several blockchain-based CTI sharing proposals have been proposed \cite{allouche2021trade} in recent years. Blockchain has a number of salient properties that can be leveraged to facilitate CTI sharing. For example, among others, blockchain's consensus, auditability, and immutability properties allow organisations to interact with one another in a distributed way without the need of a trusted third party. Another advantage of using blockchain to facilitate CTI sharing is verifiability. Considering that the information stored on the blockchain is immutable, CTI consumers are able to validate the integrity of received data and, therefore, can be sure it has not been tampered with or changed~\cite{dai2019blockchain}.

\subsection{Contributions}
\label{contributions}
Existing proposals only allow CTI producers to share intelligence in fixed way using a platform defined trust-based policy scheme. While these approaches do provide verifiable and trust-based sharing, constraining CTI producers to share in a fixed way, e.g. binary or group-based sharing, has the potential to limit the effectiveness of sharing between organisations. Subsequently, in our work we define a blockchain-based framework which provides CTI producers with the ability to share CTI \textit{differentially} using  policies/metrics which they define. Moreover, we further demonstrate how CTI producers can share in this way without compromising the verifiability of the intelligence from the CTI consumers perspective. To the best of our knowledge, our work is the first to use blockchain for trusted, verifiable, and differential CTI sharing. The major contributions of this paper are: 


\begin{itemize}
    
    
    \item We propose the concept of \textit{differential sharing} within the CTI sharing context.
    
    \item We detail how \textit{differential sharing} can provide more granuality compared to existing solutions by allowing data segmentation and \textit{pluggable} trust-based policy/metric schemes, without compromising verifiability through the use of two integrity hash schemes.  
    
    
    
    
    
    \item We discuss a detailed design for the proposed CTI sharing framework, its system functionality, and communications among the various components.
    
    \item We demonstrate and evaluate a detailed proof of concept implementation using Ethereum private blockchain. The experimental results indicate that the proposed framework is feasible and creates insignificant overheads compared to a baseline sharing scenario.
    
    \item We present a privacy and trust analysis of the proposed framework.  
    
\end{itemize}

Next, in Section~\ref{related-work}, we discuss the related work. In Section~\ref{proposed-architecture}, we introduce the proposed CTI sharing framework. Section~\ref{sytem-design} describes and evaluates the proof of concept implementation of the proposed framework and present the key results. Moreover, this section also analysis the privacy and trust of the proposed framework. Finally, Section~\ref{conclusion} concludes the paper and highlights the future work. 

\begin{table}[h]
    \centering
    \caption{Available blockchain-based CTI sharing proposals and their comparison with our work. 
    }
    \label{tab:realted_work}
    \scalebox{0.765}{
    \begin{tabular}{ccccccc}
        \toprule
            {Ref.} & {Differential} & {Sensitive} & {Producer-Defined} & {Verifiable} & {Blockchain}  \\
            {} & {Sharing} & {Data Grouping} & {Trust Scheme} & {Sharing} & {Platform} \\
        \toprule
        
        \cite{shi2022threat} & \xmark &  \xmark & \xmark & \checkmark & CITA \\ [0.5ex]
        
        \cite{badsha2020blocynfo} & \xmark & \xmark & \xmark & \checkmark & Ethereum \\ [0.5ex]
         
         \cite{allouche2021trade} & \xmark &  \xmark & \xmark & \checkmark & Ethereum \\ [0.5ex]
         
         \cite{preuveneers2020distributed} & \xmark & \xmark & \xmark & \checkmark & Hyperledger   \\ [0.5ex]
         
         \cite{nguyen2022blockchain} & \xmark & \xmark & \xmark & \checkmark & Hyperledger \\ [0.5ex]
         
         \cite{moubarak2021dissemination} & \xmark & \xmark & \xmark & \checkmark & Hyperledger \\ [0.5ex]
         
         \cite{homan2019new} & \xmark & \xmark & \xmark & \checkmark & Hyperledger \\ [0.5ex]
         
        \textbf{[Our Work]} & \checkmark & \checkmark & \checkmark & \checkmark & Ethereum \\ 
        
        \bottomrule
        
    \end{tabular}
    }
\end{table}

\section{Related Work}
\label{related-work}
In Table~\ref{tab:realted_work}, we provide a list of the available related works on blockchain-based CTI sharing and their comparison with our work. For example, 
The authors in \cite{shi2022threat} developed a blockchain-based CTI sharing platform using the CITA model. This model provides CTI producers with the ability to store sensitive data off-chain. When sharing this CTI data with authorised CTI consumers, the authors propose using point-to-point communication to ensure fine-grained access control over which CTI consumers receive the sensitive data. To ensure verifiability, a hash of the sensitive data is stored on-chain such that CTI consumers can validate the integrity of data received off-chain. While~\cite{shi2022threat} provides  trust-based verifiable sharing, CTI producers are constrained to share CTI data with consumers in a coarse-grained and binary way. 

`BloCyNfo-Share' is a blockchain-based CTI sharing framework proposed in \cite{badsha2020blocynfo}. It utilises proxy re-encryption and attribute-based encryption to achieve fine-grained access control. When sharing, CTI producers provide an encrypted copy of the data to a cloud server. When a CTI consumer requests access to this CTI, the CTI producer generates a re-encryption key and adds it to a smart contract. This re-encryption key is subsequently used by the cloud server to generate a new cipher text. The resulting cipher text is subsequently sent to the CTI consumer, who can decrypt it given they satisfy the policy specified by the CTI producer. 
Similar to \cite{shi2022threat}, in this framework, CTI producers are limited to making binary sharing decisions. Thus, unlike our approach, proposals \cite{badsha2020blocynfo} and  \cite{shi2022threat} do not provide differential sharing.

\begin{figure*}[t]
    \centering
    \includegraphics[scale=0.55]{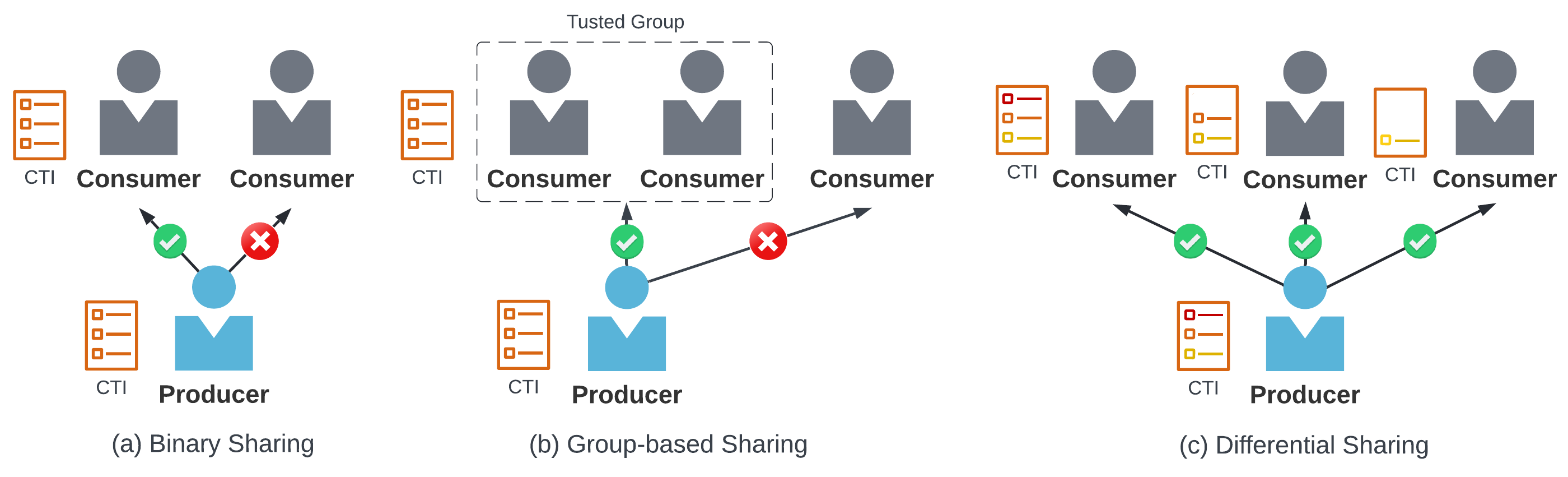}
    \caption{Illustration of our differential CTI sharing compared to traditional binary and group-based CTI sharing approaches.}
    \label{fig:turst-group-df}
\end{figure*}

In \cite{allouche2021trade}, the blockchain-based CTI sharing platform TRusted Anonymous Data Exchange (TRADE) is proposed. To facilitate the exchange of CTI data, the TRADE platform proposes the use of off-chain data storage, policy-based access control, on-chain policy validation and off-chain data exchange using Trusted Automated Exchange of Intelligence Information (TAXII)~\cite{jordan_varner_2021}. When a CTI producer wants to share, 
they add an integrity hash and access policy (e.g., eXtensible Access Control Markup Language (XACML) policy) to the blockchain. 
However, TRADE again limits CTI producers to binary sharing decisions. 

In contrast to \cite{badsha2020blocynfo}, proposal \cite{preuveneers2020distributed} utilises the existing CTI sharing platform MISP \cite{wagner2016misp} to facilitate the exchange of CTI between organisations. To ensure that the privacy of sensitive data is maintained, a \textit{Trustworthy API for Threat Intelligence Sharing} (TATIS) Reverse Proxy is used. This proxy uses ciphertext attribute-based encryption to restrict the access of sensitive data to trusted organisations. To provide verifiability, CTI producers store a hash of the original data on-chain. However, like the above proposals, \cite{preuveneers2020distributed} also limits producers to binary sharing decisions.

In a few other proposals, e.g., \cite{nguyen2022blockchain},  \cite{moubarak2021dissemination}, and \cite{homan2019new}, a group-based approach for sensitive CTI sharing is utilised. Instead of limiting CTI producers to binary sharing decisions, a group-based approach allows CTI producers and consumers to exchange sensitive CTI as part of a trusted group. Moreover, these proposals utilise the Traffic Light Protocol (TLP). Instead of limiting CTI producers to share in a binary way, more or less sensitive data can be shared using a layered approach. In this case, layered refers to a grouping of organisations into trust groups (e.g., red, amber, green, and white), in which CTI of varying sensitivity can be shared. 
However, the major limitation of a group-based approach is that it assumes organisations trust each other in a pre-defined and fixed way. While these approaches extend sharing beyond binary decisions, we argue they still lack the granularity required for more effective CTI sharing. Subsequently, our proposal allows CTI producers to share differentially, and therefore supports dynamic and time-varying trust relationships.  


In summary, the existing blockchain-based CTI sharing proposals limit a CTI producer's ability to share sensitive CTI in a coarse-grained manner, either based on binary share/deny policies or as part of a pre-defined trust group. In most cases, the existing solutions provide trust-based policies/metrics that limit the dissemination of sensitive data to trusted organisations, but they fail to deliver more granular sharing. Given that a more fine-grained sharing mechanism is required to facilitate effective CTI sharing, we propose a framework with the concept of differential sharing discussed in the next section. 

\section{Proposed Framework}
\label{proposed-architecture}


In this section, we present the proposed blockchain-based CTI sharing framework. First, we discuss the concept of differential CTI sharing in detail. Then we present the various components of the architecture. We also discuss the system functionality and communication among these components. 

\subsection{Differential Sharing}
\label{differential}
In Section~\ref{introduction}, we highlight that organisations are likely to differ in their perceptions of data sensitivity and what trust-based policy/metrics should be used to determine a CTI consumers access. We propose the concept of \textit{differential} CTI sharing, which aims to address this challenge.  Figure~\ref{fig:turst-group-df} compares differential CTI sharing with binary and trusted group-based CTI sharing. From this comparison, we highlight that differential sharing can tailor the CTI data to be shared with each CTI consumer. As a result, the proposed \textit{differential} sharing approach does not restrict sharing to only highly trusted CTI consumers, and therefore, allows other CTI consumers to benefit from CTI data they otherwise would not have received.

\begin{figure} [ht]
\includegraphics[scale=0.69]{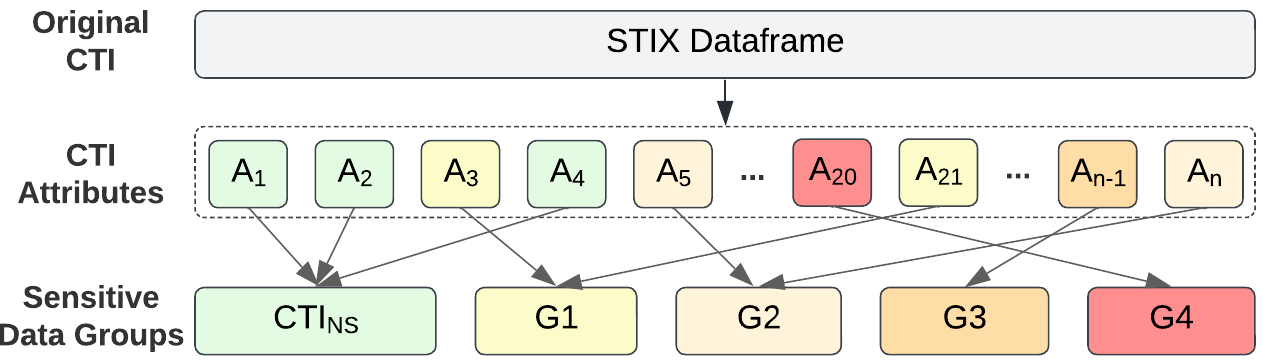}
\caption{Sensitive data groups based on CTI attributes.}
\label{fig:stix-cti}
\end{figure}

Given that CTI data contains many attributes, e.g., \textit{chained objects} in Structured Threat Information Expression (STIX), their individual sensitivity can be evaluated by the CTI producer. Furthermore, as these attributes are likely to vary in their sensitivity, attributes of similar sensitivities can be grouped together. For example, \textit{chain-objects} of type vulnerability often contain specific software versions that could be valuable information to potential attacks, while \textit{chain-objects} of type indicator often only contain IoC. As a result, based on the information contained within specific STIX \textit{chain-objects}, Fig.~\ref{fig:stix-cti} demonstrates how CTI attributes can be broken down into a variable number of sensitive data groups. 


\begin{figure} [t]
\centering
\includegraphics[scale=0.82]{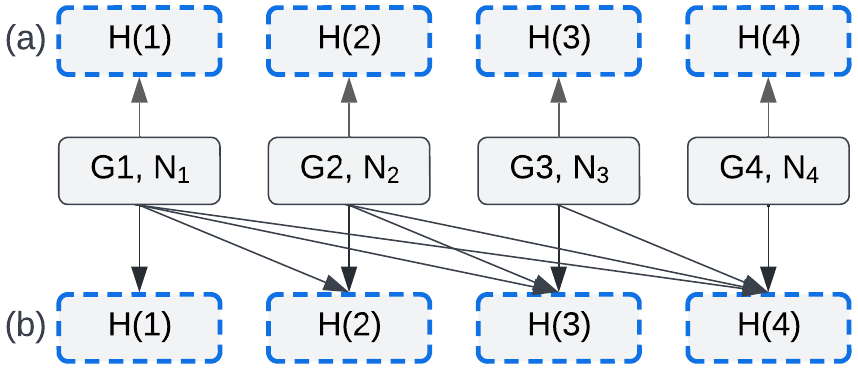}
\caption{Proposed single hash (a) and multi hash (b) approaches. Note, dotted boxes represent integrity hashes. ($N_i$ = One-time random nonce).} 
\label{fig:intgrity-hashes}
\end{figure}

Once the CTI has been segmented using the above method, a subset of these sensitive data groups can be given to CTI consumers based on a single or set of trust-based policies/metrics (e.g., XACML policy). By allowing CTI producers to segment CTI data into groups based on a sensitivity assessment, a variable proportion of these groups can be shared with each CTI consumer. To ensure verifiability of the received data, we propose the use of two group-based hashing approaches: single hash and multi-hash. Given that CTI consumers receive a distinct number of the sensitive data groups, a set of integrity hashes constructed using either of these two approaches is required. In Fig.~\ref{fig:intgrity-hashes} we illustrate how these group-based hashes can be constructed using a single and multi hash approach. 


Sharing CTI using the proposed \textit{differential} approach gives CTI producers the ability to share with more granularity compared to traditional solutions, e.g., binary and group-based. We further note that greater granularity is achieved in two ways: (i) CTI producers are given the ability to define what subset of the CTI data is shared with an individual or group of CTI consumers, and (ii) CTI producers are also able to change how they segment CTI data and the policies/metrics they use to evaluate trust each time they share.

\subsection{Architectural Components}
\label{components}
In Fig.~\ref{fig:architecture}, we illustrate the proposed blockchain-based CTI sharing framework. It contains the following components:

\begin{itemize}
    \item \textbf{Organisation:} An entity that participates in CTI sharing. Organisations can assume the role of a producer and/or consumer. Organisations can participate in both roles over time. However, we assume that an organisation performs only one role at a time for a specific interaction.
    
    \item \textbf{Request Manager:} A component within an organisation responsible for handling and submitting access requests of CTI data. The request manager role is primarily broken into two tasks, \textit{submitting} an access request and \textit{processing} an access request. They are as follows:  
    \begin{enumerate}
        \item Submitting an access request occurs when new CTI is made available. Given that another organisation has added a CTI, the request manager submits an access request to the blockchain. 
        
        \item When an organisation adds a new CTI, the organisation's request manager is responsible for processing the access requests added by other organisations.
    \end{enumerate}
    
    
    
    
    \item \textbf{Data Manager:} A component inside of an organisation responsible for storing sensitive CTI data and their corresponding access policies. This component is also responsible for indicating to other organisations that new CTI is available. Subsequently, it adds content on and off-chain when an organisation acts as a CTI producer.
    
    \item \textbf{Access Manager:} A component inside an organisation responsible for determining what portion of requested CTI should be shared with a CTI consumer. Using the access policies stored by the data manager and credentials (e.g., identity-proof) provided by the CTI consumer, the access manager determines what portion of the CTI data should be given to a particular CTI consumer.

\begin{figure}[t]
\centering
\includegraphics[scale=0.755]{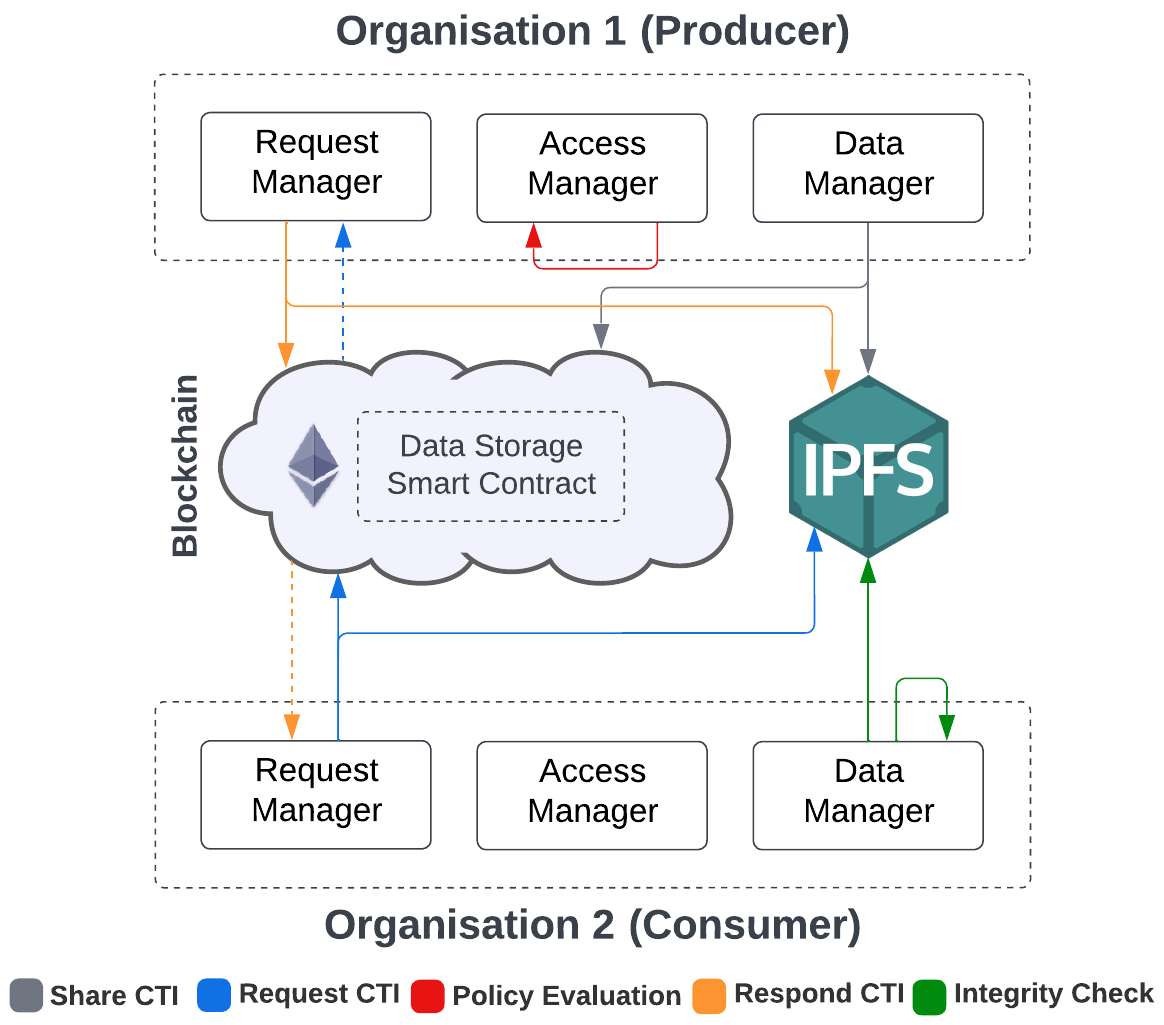}
\caption{Proposed blockchain-based CTI sharing framework.}
\label{fig:architecture}
\end{figure}

    \item \textbf{Blockchain:} A distributed digital ledger of information that is shared, duplicated and distributed across the entire network of computer systems. It can be seen as a shared database that stores a series of digital records using cryptographic operations (called hash). 
    Ethereum blockchain is used for our proof-of-concept implementation. 
    
    \item \textbf{Smart Contract:} 
    Self-executable programs stored on a blockchain. We introduce a \textit{data storage} smart contract in our framework. It is composed of three functions: (i) \textit{share} function provided by the smart contract is used by a CTI producer to indicate the availability of CTI, (ii) \textit{request} function provided by the smart contract is used by a CTI consumer to request access to sensitive CTI data, and (iii) \textit{response} function provided by the smart contract is used by a CTI producer to respond to a request submitted to the \textit{request} function.
    
    
    
    \item \textbf{Interplanetary File System (IPFS):} A peer-to-peer hypermedia protocol that is designed to distribute the storage of data across a network of computers. IPFS provides highly available and consistent data storage of large files that otherwise would be expensive to store using a traditional blockchain network (e.g., Ethereum). IPFS is utilised to store off-chain data associated with the sharing process \cite{benet2014ipfs}. 
    
    
    
\end{itemize}



\subsection{System Functionality}
\label{system-functionality}

This section provides an overview of the proposed CTI sharing framework's functionality. The framework follows five main phases to share and consume CTI. They are: 

\underline{{\textbf{\textit{Share CTI:}}}} The objective of this phase is for the CTI producer to create and share all of the data 
required to facilitate the exchange of CTI. Given that an organisation has produced CTI data, a sensitivity analysis is conducted by the CTI producer. This analysis identifies sensitive CTI attributes and segments them into a number of sensitivity groups (cf. Fig.~\ref{fig:stix-cti}). 






Once the sensitive data groups have been defined, a set of \textit{integrity hashes} are created. To construct these hashes, the single or multi hash approach (cf. Fig.~\ref{fig:intgrity-hashes}) can be used by the CTI producer. For both of the proposed approaches, the sensitive data groups and a set of one-time random nonce's are used to construct the integrity hashes. In this case, the set of one-time random nonce's are used to protect the privacy of sensitive data (more discussion in Section~\ref{privacy-trust}). To generate these integrity hashes, a single or set of data groups are hashed using these nonce's in sequence. 

As discussed in Section~\ref{related-work}, several trust-based metric/policy schemes have been presented by existing works. Subsequently, our framework allows CTI producers to plug-in a scheme dynamically. For example, an organisation might assign an individual XACML policy for each of the sensitive data groups. 

Once the above process has been completed, the sensitive data groups and access policy are stored securely off-chain by the CTI producer. Moreover the non-sensitive CTI data group, set of integrity hashes and access policy are added to IPFS, while its associated content identifier (CID) (IPFS File Pointer) and CTI metadata are added to the \textit{data storage} smart contract. 


\underline{{\textbf{\textit{Request CTI:}}}} During this phase, a CTI consumer requests access to a particular CTI of interest. An access request is submitted via a \textit{data storage} smart contract. This smart contract is the same one used during the \textit{Share CTI} phase. The access request indicates which CTI data the CTI consumer wishes to access and contains the credentials (e.g., identity-proof) required to meet the associated access policy.

To ensure that these credentials are shared in a privacy-preserving way, they are encrypted using the CTI producer's public key before being added to IPFS. By encrypting their credentials, a CTI consumer can ensure that only the intended CTI producer can access these them. Once the access request has been submitted, the CTI producer receives the request, accesses the encrypted credentials from IPFS using the provided IPFS CID, and decrypts them using their private key. 

\underline{{\textbf{\textit{Policy Evaluation:}}}} During this phase, the CTI producer processes the access request submitted by the CTI consumer. To determine the sensitive data groups that the CTI consumer should have access to, the set of credentials provided during the \textit{Request CTI} phase and the access policy associated with the requested CTI is used. Once the provided credentials are validated, the CTI producer evaluates the access policy to determine which subset of the sensitive data groups the requesting CTI consumer should receive.



\underline{{\textbf{\textit{Response CTI:}}}} During this phase, the CTI producer provides the CTI consumer with the sensitive data groups they are able to access. Note, in the \textit{Policy Evaluation} phase above, the CTI producer determines which subset of the sensitive data groups a CTI consumer is able to access. To ensure that the CTI consumer can validate the integrity of the provided sensitive data groups, the corresponding one-time random nonce values are also included in the response. To ensure that the response data is exchanged in a privacy-preserving way, it is encrypted using the CTI consumer's public key. Once encrypted, the response data is added to IPFS and the associated IPFS CID is added to the \textit{data storage} smart contract by the CTI producer. The CTI consumer can then access the encrypted response data using the provided IPFS CID. To decrypt the response data, the CTI consumers uses their private key. 

\underline{{\textbf{\textit{Validate CTI:}}}} During this phase, the CTI consumer validates the integrity of the received sensitive data. As part of the \textit{Share CTI} phase, the CTI producer added a set of integrity hashes to IPFS. Subsequently, the CTI consumer can access these integrity hashes and use them to validate the sensitive CTI data received during the \textit{Response CTI} phase. 

\begin{figure}
\centering
\includegraphics[scale=.48]{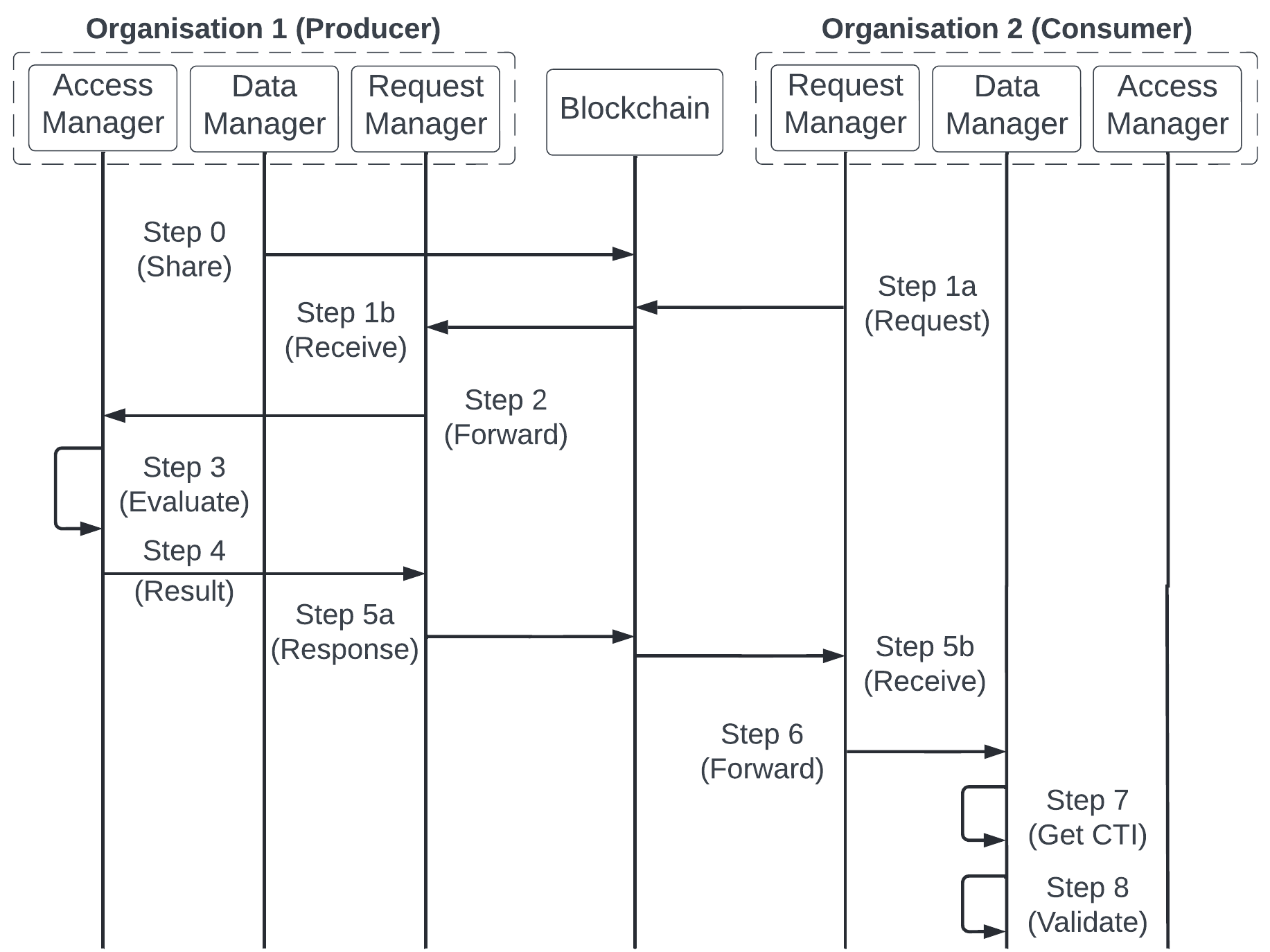}
\caption{Communication between various components of the proposed CTI sharing framework (on-chain interactionss).}
\label{fig:interactions}
\end{figure}

\subsection{Communications}
\label{communication}


In Fig.~\ref{fig:interactions}, we illustrate the communication flow within our CTI framework. Note that for simplicity, we contemplate the interactions of various components with the blockchain. Therefore, only on-chain interactions are shown, and interactions with IPFS are excluded. The steps are as follows: 

\textbf{Step 0:} CTI producer completes the initial setup steps. Firstly, the CTI data is segmented into a variable number of sensitivity data groups (cf. Fig.\ref{fig:stix-cti}). The sensitive data groups and a set of one-time random nonces are used to construct a set of integrity hashes using either of the proposed methods (cf. Fig.\ref{fig:intgrity-hashes}). The non sensitive data group, CTI producer-defined access policy, and integrity hashes are added to IPFS. The access policy and sensitive data groups are stored off-chain by the CTI producer's \textit{access manager}. Lastly, the IPFS CID and CTI metadata (e.g., threat type) are added to the \textit{data storage} smart contract using the \textit{share} function.


\textbf{Step 1a:} The CTI consumer's \textit{request manager} adds an access request to the \textit{data storage} smart contract using the \textit{request} function. As discussed in Section~\ref{system-functionality}, the request contains a IPFS CID that is associated with an encrypted set of the CTI consumers' credentials. 


\textbf{Step 1b:} The CTI producer's \textit{request manager} receives the access request. The \textit{request manager} decrypts the credentials associated with the IPFS CID provided by the CTI consumer using their private key.  


\textbf{Step 2:} The CTI producer's \textit{request manager} forwards the decrypted credentials and requested CTI to the \textit{access manager}. 

\textbf{Step 3:} The CTI producers \textit{access manager} validates the received credentials (e.g., certificate signature is checked). Using the access policy associated with the requested CTI, and the verified credentials, the \textit{access manager} assess the subset of sensitive data that should be provided to the CTI consumer.


\textbf{Step 4:} The CTI producers \textit{access manager} returns the accessible sensitive data groups and their associated one-time random nonce to the CTI producers \textit{request manager}. 


\textbf{Step 5a:} The CTI producer's \textit{request manager} responds to the access request submitted during \textit{Step 1a} using the \textit{data storage} smart contract's \textit{response} function. As discussed in Section~\ref{system-functionality}, this response contains an IPFS CID that is associated with the encrypted set of the accessible sensitive data groups and one-time random nonce's.


\textbf{Step 5b:} The CTI consumer's \textit{request manager} receives the response. The \textit{request manager} decrypts the accessible sensitive data groups and one-time random nonce's associated with the IPFS CID using their private key.   


\textbf{Step 6:} The CTI consumer's \textit{request manager} forwards the received sensitive data, one-time random nonce's and original IPFS CID referencing the original CTI data, to the CTI consumers \textit{data manager}.


\textbf{Step 7:} Using the original IPFS CID, the CTI consumers \textit{data manager} gets the set of integrity hashes shared by the CTI producer during \textit{Step 0}. 

\textbf{Step 8:} Using the received sensitive data groups and one-time random nonce's, the CTI consumer's \textit{data manager} computes the integrity hashes. This hash(s) is then compared to the set of integrity hashes received during \textit{Step 7}.

\section{System Design and Evaluation}
\label{sytem-design}
This section discusses the
implementation of the proposed CTI sharing framework. We evaluate the developed proof-of-concept prototype using several key on- and off-chain metrics, e.g. \textit{gas cost} and \textit{computational performance}. In addition, we also provide a privacy and trust analysis.  

\begin{figure}[t]
    \centering
    \includegraphics[scale=0.42]{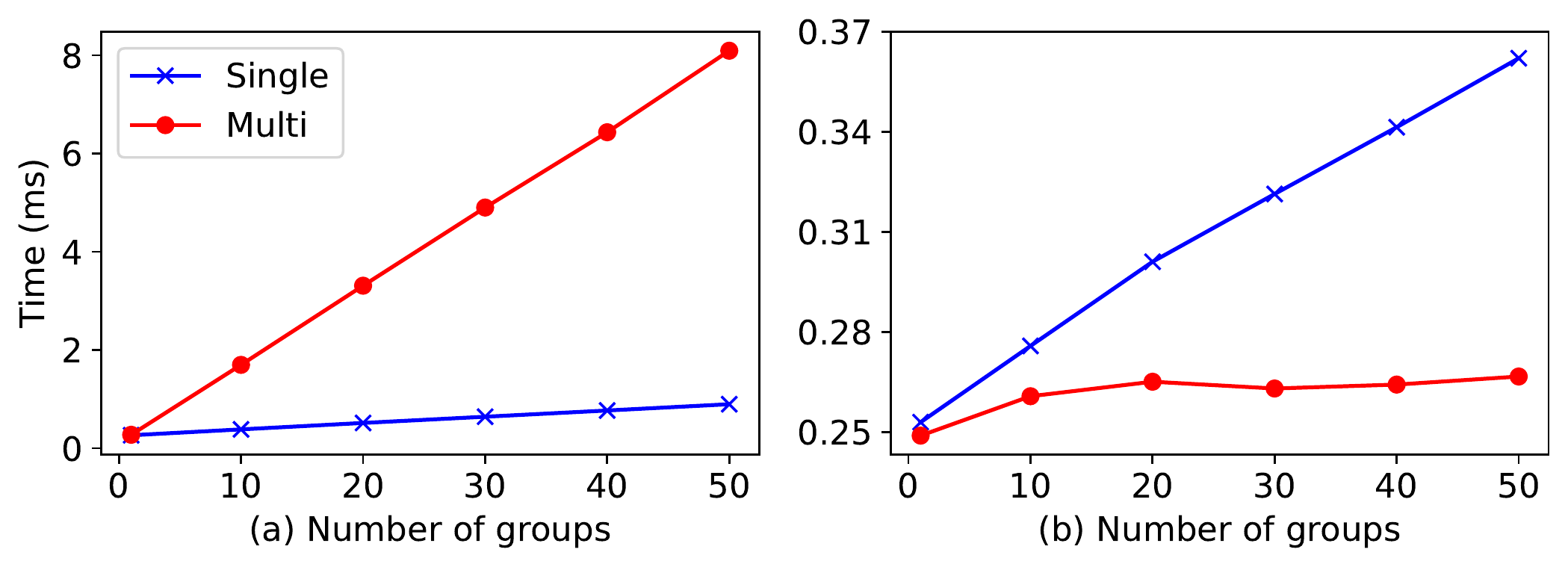}
    \caption{Sample data integrity hash (a) \textit{generation}, and (b) \textit{validation} time (ms) using the single (Single) and multi (Multi) hash approaches.}
    \label{fig:sample_both}
\end{figure}


\subsection{Implementation}
\label{implementation}
We implemented a proof-of-concept of our proposed framework using Ethereum blockchain, in which we deploy our \textit{data storage} smart contract written in Solidity v0.8.11. We select the Ethereum platform as it can be deployed as a public or private blockchain, allowing higher flexibility when evaluating our framework. We wrote a Python script using \texttt{web3.py} library \cite{web3-lib} to evaluate the blockchain implementation.

In addition, we developed the off-chain components of our framework in the Python programming language (e.g., request, access, and data managers). Specifically, we used \texttt{hashlib} library \cite{hash-lib} to implement the proposed integrity hashing mechanisms (cf. Fig.~\ref{fig:intgrity-hashes}), where we use the SHA-256 hashing function and a public CTI dataset available on GitHub~\cite{STIX-GITHUB} to evaluate its performance.
In addition, we also generate a number of synthetic data files larger and smaller than the CTI sample data. For both the sample and synthetic files, the data is segmented into various groups. We deployed a resource-constrained Amazon Elastic Compute Cloud instance (i.e., \texttt{t2.micro}) to benchmark the off-chain operations, as CTI sharing may also occur in settings with computational limitations (e.g., in an IoT setting). As such, we consider these constrained environments as a \textit{worst case} scenario when measuring the effect of additional hashing operations.

\subsection{Results}
\label{results}

To evaluate the on-chain components of the developed \textit{data storage} smart contract, we evaluated the transaction cost associated with the \textit{share}, \textit{request} and \textit{response} functions under two different market conditions, \textit{current} (1 gas = 20 gwei, 1 Eth = \$1,097.2) and all time \textit{high} (1 gas = 979 gwei, 1 Eth = \$4891.70). In this context, gas cost represents the execution cost of the smart contact that is payed for in Eth (i.e., a gas fee that is required for an Ethereum blockchain network user to conduct a transaction on the network) based on the cost per gas unit, measured in gwei (note, each gwei is equal to $10^{-9}$ Eth).

In Table~\ref{tab:cost}, the gas cost associated with the \textit{data storage} smart contract functions are calculated using the two above-defined market conditions. These results show that under \textit{current} market conditions, the transaction cost associated with sharing and requesting CTI is US\$0.96 and US\$2.57 respectively. As a result, the additional overhead created by our proposed differential sharing mechanism is US\$2.57. Given that this increased cost is incurred by CTI consumer who in-exchange receive potentially valuable threat information, we argue this additional cost is unlikely to prevent participating.

Furthermore, Table~\ref{tab:cost} \textit{High (US\$)}, also highlights that transaction costs increase drastically (21762\%) in the case where the market price for both gas and Eth rise significantly. While extreme, this result highlights the potential effect fluctuations in transaction costs may have on the effectiveness of the proposed framework. Subsequently, we argue that a private Ethereum network should be considered in a real world context, as this would allow participating organisations to better moderate these costs.    

Additionally, we also evaluated the off-chain computational overheads associated with the proposed integrity hashing schemes (cf. Fig.~\ref{fig:intgrity-hashes}) using the sample and synthetic data. For each of these data types, the execution time associated with the generation and validation of an integrity hash set using both of the proposed hashing schemes was measured. In addition, the execution time associated with hashing the sample and synthetic data was also measured as a baseline. Given that the generation and validation execution times are highly dependent on the number of additional hashing operations, the baseline case indicates the additional overheads created by the two proposed hashing schemes relative to the size of the input data. As discussed in Section \ref{implementation}, these operations were performed on an Amazon Elastic Compute Cloud \texttt{t2.micro} instance, with the average of 1000 iterations recorded.

\begin{figure}[t]
    \centering
    \includegraphics[scale=0.38]{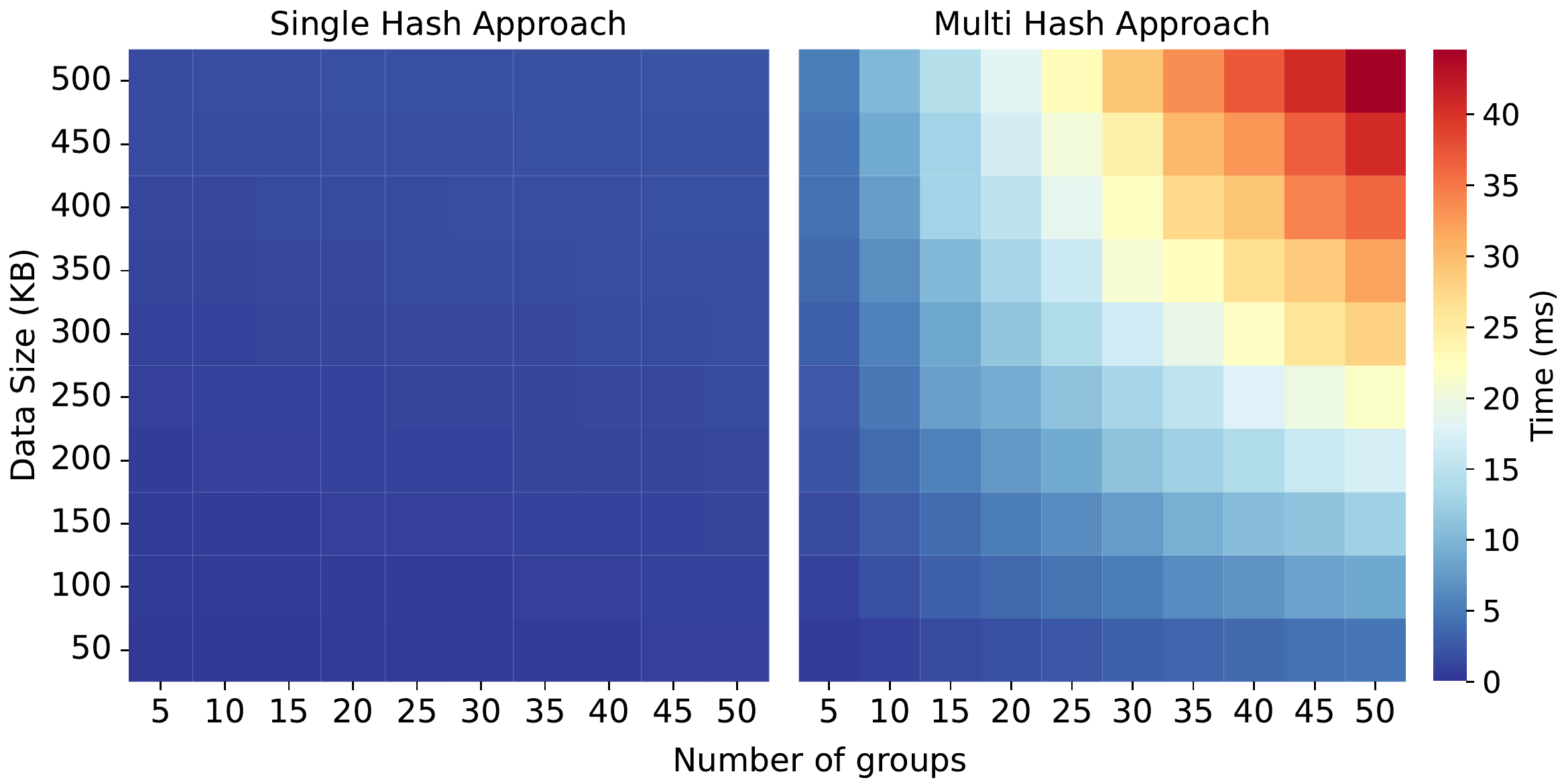}
    \caption{Synthetic data integrity hash \textit{generation} time using the single (Single) and multi (Multi) hash approaches.}
    \label{fig:datasize_gen}
\end{figure}

\begin{figure}[t]
    \centering
    \includegraphics[scale=0.38]{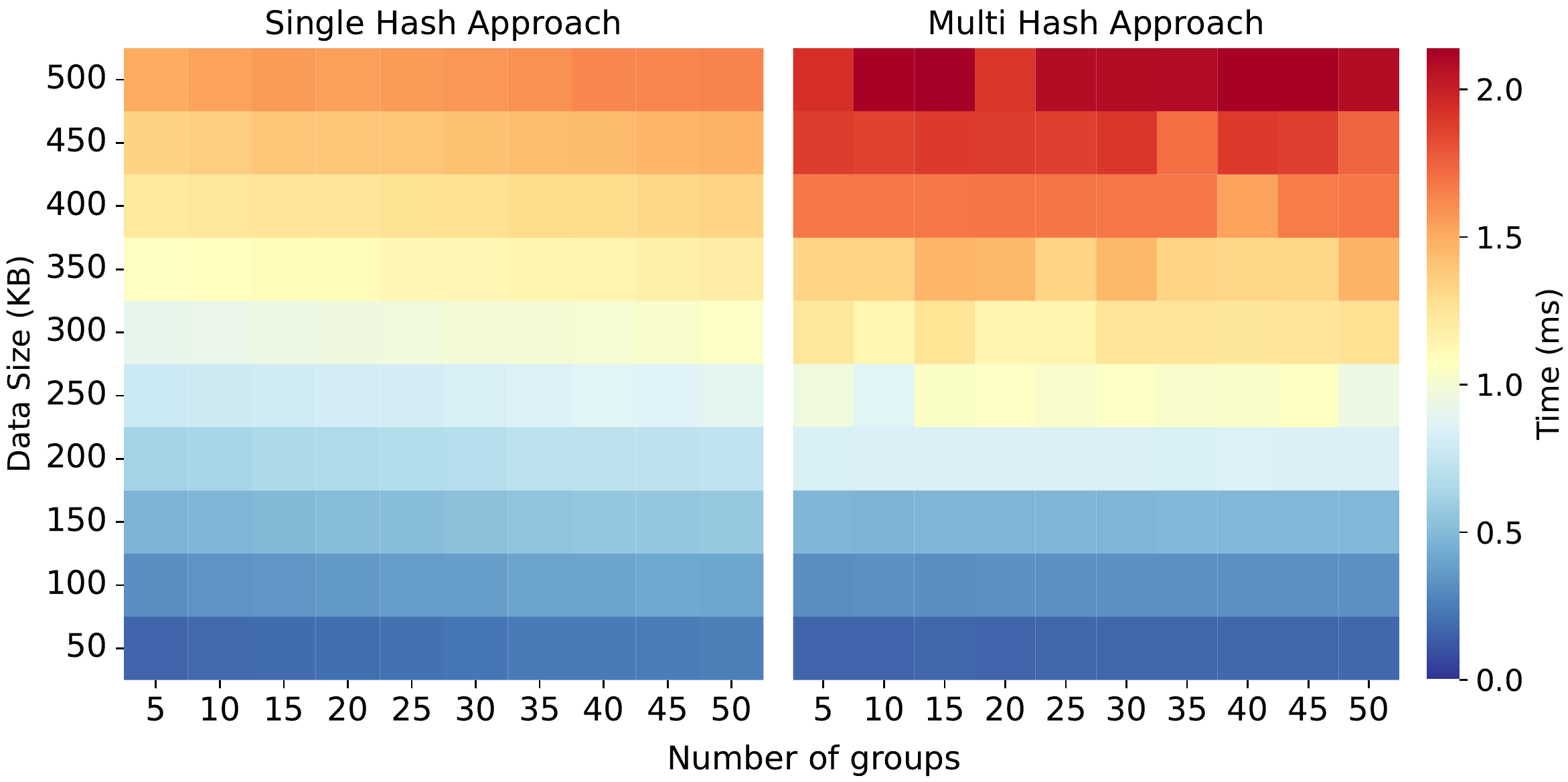}
    \caption{Synthetic data integrity hash \textit{validation} time using the single (Single) and multi (Multi) hash approaches.}
    \label{fig:datasize_val}
\end{figure}

In Table~\ref{tab:compute_results}, a summary of the key results presented in Figs.~\ref{fig:sample_both}-\ref{fig:datasize_val} is provided. These results show that both the single hash and multi hash approaches result in an increased execution time compared to the baseline. However, we argue that these additional overheads are outweighed by the broader access to CTI and the greater granularity achieved by the proposed \textit{differential} approach. Moreover, given that these results were obtained using a resource-constrained test bed, they indicate that the proposed \textit{differential} sharing mechanism can operate efficiently in resource-constrained scenarios, e.g., with IoT.

When the performance of these two approaches (i.e, single and multi hash) using the sample data are compared (cf. Fig.~\ref{fig:sample_both}), two trends can be observed. In terms of generating the integrity hashes, the multi hash approach can be observed to have an increased execution time when compared to the single hash approach. This result is  due to the redundancy associated with hashing the same data exponentially more times as the number of data groups get larger for the multi hash case (cf. Fig.~\ref{fig:intgrity-hashes}). In contrast, the opposite relationship can be observed if the validation times of these two approaches are compared. As the number of data groups increases, it can be observed that the single hash approach increases in execution time compared to the multi hash approach. This result is due to the redundancy associated with the number of comparisons required to validate a set of groups being equal to number of groups for the single hash case (cf. Fig.~\ref{fig:intgrity-hashes}). 

For the synthetic data, similar trends were also observed. In Figs.~\ref{fig:datasize_gen} and \ref{fig:datasize_val}, the single hash generation and multi hash validation efficiency differences are mostly consistent. However, we observed that the multi hash validation efficiency reduced proportional to the size of synthetic data. More specifically, after a data size of 150(KB), we observed that the single hash approach was either comparable or more efficient in terms of validation performance. 

Based on both the sample and synthetic data results, we argue that the multi-hash approach is more efficient. Given that CTI data is shared with \textit{C} number of CTI consumers, integrity hash validation can be considered a \textit{C}-time operation, and therefore, additional validation overheads are exacerbated in a real-world setting where \textit{C} is denoted as a large number. In contrast, integrity hash generation is a one-time operation completed by the CTI producer each time they share. However, we note that in circumstances where CTI data is significantly larger than the sample data, the single hash approach may become more efficient, considering the validation time results for the 150-500KB synthetic data in Fig.~\ref{fig:datasize_val}. 

Using the discussed results, we make three major findings: (i) the proposed differential sharing mechanism does not significantly increase transaction costs, (ii) market conditions could affect the effectiveness of CTI sharing, and (iii) the proposed integrity hashing mechanisms (cf. Fig.~\ref{fig:intgrity-hashes}) do not create large computational overheads, with the multi hash approach being more efficient for CTI data size less than 150KB.

\begin{table}[]
    \centering
    \caption{Gas cost and price in US\$ for each \textit{data storage} smart contract function.}
    \label{tab:cost}
    \scalebox{1}{
    \begin{tabular}{cccc}
        \toprule 
        Function & Gas Cost & Current (US\$) & High (US\$)  \\
        \toprule 
        Share & 43,897 & \$0.96 & \$210.22 \\ [0.5ex]
        Request & 66,628 & \$1.46 & \$319.07 \\ [0.5ex]
        Response & 50,625 & \$1.11 & \$242.44 \\ [0.5ex]
        \textbf{Total} & \textbf{161,150} & \textbf{\$3.53} & \textbf{\$771.73} \\
        
        \bottomrule
    
    \end{tabular}
    }
\end{table}

\begin{table}[]
    \centering
    \caption{Summary of computational times associated with Figs.~\ref{fig:sample_both}-\ref{fig:datasize_val}. (Gen = Generation Time, Val = Validation Time, SH = Single Hash, MH = Multi Hash).}
    \scalebox{0.94}{
    \begin{tabular}{cccccccc}
        \toprule
   
        Data & Size & Baseline & Number of & \multicolumn{2}{c}{Gen (ms)} & \multicolumn{2}{c}{Val (ms)} \\
        {Type} & {(KB)} & {(ms)} & {Group} & SH & MH & SH & MH \\
        \toprule      
        \multirow{3}{*}{Sample} & 
            \multirow{3}{*}{94} & 
                \multirow{3}{*}{0.24} &
                10 & 0.38 & 1.69 & 0.27 & 0.26 \\ [0.5ex]
                & & & 20 & 0.51 & 3.30 & 0.30 & 0.27  \\ [0.5ex]
                & & & 50 & 0.89 & 8.10 & 0.36 & 0.26  \\ 
                
        \midrule
        
        \multirow{9}{*}{Synthetic} & 
                
            \multirow{3}{*}{50} & 
                \multirow{3}{*}{0.14} &
                5 & 0.22 & 0.54 & 0.16 & 0.16   \\ [0.5ex]
                & & & 20 & 0.41 & 1.9 & 0.2 & 0.17  \\ [0.5ex]
                & & & 50 & 0.78 & 4.6 & 0.26 & 0.17  \\ [0.5ex]
              \cline{2-8} 
            & \multirow{3}{*}{200} & 
                \multirow{3}{*}{0.55} &
                5 & 0.67 & 2.1 & 0.61 & 0.83  \\ [0.5ex]
                & & & 20 & 0.88 & 7.2 & 0.66 & 0.83  \\ [0.5ex]
                & & & 50 & 1.2 & 17 & 0.72 & 0.83  \\  [0.5ex]
                  \cline{2-8} 
            & \multirow{3}{*}{500} & 
                \multirow{3}{*}{1.4} &
                5 & 1.6 & 5.1 & 2.5 & 1.9  \\ [0.5ex]
                & & & 20 & 1.8 & 18 & 1.5 & 1.9  \\ [0.5ex]
                & & & 50 & 2.2 & 45 & 1.6 & 2.1  \\ 
        \bottomrule
    \end{tabular}}
    \label{tab:compute_results}
\end{table}

\subsection{Privacy and Trust Analysis}
\label{privacy-trust}
Now we examine a number of privacy and trust-based considerations with our proposed framework. From a CTI producer's perspective, it is important that sensitive data is shared in a privacy-preserving way. Subsequently, the proposed differential approach seeks to provide a more granular approach that maintains privacy at rest and in transit. To ensure that privacy is maintained at rest, sensitive data groups are stored securely off-chain by CTI producers, and therefore cannot be accessed by untrusted entities. Moreover, when sensitive data groups are exchanged with trusted CTI consumers, encryption is used to ensure that \textit{man-in-the-middle} and \textit{eavesdropping} attacks (e.g., sniffing or snooping) are not possible. Finally, the use of one-time random nonce's as part of the integrity hash generation and validation process, ensures that the privacy of sensitive data common across multiple sharing instances or which contains low entropy (e.g., software versions), are protected from dictionary and brute force attacks. 


Further, two major trust-based factors must also be considered. Firstly, CTI producers must be provided with a dynamic approach to trust management that ensures that sensitive data is only shared with trusted entities. Subsequently, our framework allows CTI producers to define a trust-based policy/metric scheme that best suits their requirements. Finally, the verifiability of CTI data from the perspective of CTI consumers must also be considered. For say, if a CTI consumer receives a subset of the CTI data, it could be possible for a malicious CTI producer to modify this data depending on the identity of the requesting organisation (e.g., a competitor). In such a case, to protect CTI consumers from this attack, our proposed integrity hashing schemes ensure that CTI consumers can validate that the received subset of the CTI data is contained within the original set of CTI data shared by the CTI producer. 

\section{Conclusion}
\label{conclusion}

We have proposed a blockchain-based CTI sharing framework that allows for the trusted, verifiable, and differential exchange of CTI between producers and consumers. We refer to differential as the ability of a CTI producer to control the amount of information exchanged with a CTI consumer. The framework thus allows CTI producers to segment CTI data into sensitive data groups, which can be variably shared with CTI consumers in a differential manner. 
Our results showed that the proposed framework could facilitate the exchange of CTI between producers and consumers in a differential way without compromising trust or variability. An interesting direction for future work is to conduct a more comprehensive study to examine the impact of incentives on CTI sharing. 

\section*{Acknowledgment}
The authors acknowledge the support of the Commonwealth of Australia and Cybersecurity Research Centre Limited.

\ifCLASSOPTIONcaptionsoff
  \newpage
\fi

\bibliographystyle{IEEEtran}
\bibliography{bare-jrnl}

\end{document}